\documentclass[11pt,dvips]{article}
cm \voffset = -2truecm

\usepackage{graphicx}
\usepackage{amssymb}
\usepackage{latexsym}

\begin{document}

\begin{center}

%%% \title{Generalized Coherent States for Polynomial Weyl-Heisenberg Algebras}

{\bf {\large GENERALIZED COHERENT STATES FOR \\ 
POLYNOMIAL WEYL-HEISENBERG ALGEBRAS}}
%%%%%%%%%%%%%%%%%%%
\vspace{2cm}

{\bf Maurice R. Kibler$^{1,2,3}$ and Mohammed Daoud$^{1,2,3,4}$}

\vspace{0.8cm}

\noindent$^1$ Universit\'e de Lyon, 69361 Lyon, France \\
$^2$ Universit\'e Claude Bernard Lyon 1, 69622 Villeurbanne, France \\
$^3$ CNRS/IN2P3, Institut de Physique Nucl\'eaire, 69622 Villeurbanne, France \\
$^4$ D\'epartement de Physique, Facult\'e des Sciences, Agadir,
Morocco

\vspace{0.5cm} \noindent E-mail : m.kibler@ipnl.in2p3.fr and m\_daoud@hotmail.com

\vspace{1.5cm}
%%%%%%%%%%%%%%%%%%%

\end{center}

\begin{abstract}
\noindent It is the aim of this paper to show how to construct {\it \`a la} Perelomov and {\it \`a la} 
Barut-Girardello coherent states for a polynomial Weyl-Heisenberg algebra. This algebra
depends on $r$ parameters. For some special values of the parameter corresponding to
$r = 1$, the algebra covers the cases of the su(1,1) algebra, the su(2) algebra and the
ordinary Weyl-Heisenberg or oscillator algebra. For $r$ arbitrary, the generalized Weyl-
Heisenberg algebra admits finite or infinite-dimensional representations depending
on the values of the parameters. Coherent states of the Perelomov type are derived
in finite and infinite dimensions through a Fock-Bargmann approach based on the use
of complex variables. The same approach is applied for deriving coherent states of
the Barut-Girardello type in infinite dimension. In contrast, the construction of {\it \`a la}
Barut-Girardello coherent states in finite dimension can be achieved solely at the
price to replace complex variables by generalized Grassmann variables. Finally, some preliminary 
developments are given for the study of Bargmann functions associated with some of the 
coherent states obtained in this work. 

\vspace{3.5cm}


\noindent keywords: polynomial Weyl-Heisenberg algebra; su(2); su(1,1); oscillator algebra; generalized 
coherent states; Barut-Girardello coherent states; Perelomov coherent states; Bargmann functions
\end{abstract}

\vspace{1.5cm}

\newpage

\section{INTRODUCTION}

Coherent states are of paramount importance in physics (e.g., in 
quantum optics and quantum information theory) and mathematical 
physics (e.g., in probability theory, applied group theory, path integral formalism 
and theory of analytic functions) [1-6]. They are generally 
associated with a quantum system (like the oscillator or the 
Morse or the P\"oschl-Teller systems) or an algebra (like the 
Weyl-Heisenberg algebra or a Lie algebra). The most well-known 
coherent states concern the harmonic oscillator system [7]. The 
coherent states for the su$(2)$ and su$(1,1)$ algebras play also 
an important role in various fields of theoretical and mathematical 
physics since the pioneer works by Barut and Girardello [8] and by 
Perelomov [1]. In recent years, as extensions of these well-known 
examples, generalized coherent states were the object of numerous 
studies (see for instance [9-16]).  

It is the object of the present article to report on a new 
construction, based on a {\it \`a la} Fock-Bargmann approach, 
of generalized coherent states associated with a polynomial 
Weyl-Heisenberg algebra. The construction is achieved both in 
finite and infinite dimensions. 

The paper is organized as follows. In Section 2, basics about the coherent 
states for the harmonic oscillator are briefly reviewed in order to understand 
which results can or cannot be generalized. Section 3 deals with the study of a 
polynomial Weyl-Heisenberg algebra which is an extension of the Weyl-Heisenberg 
algebra for the one-dimensional harmonic oscillator. The main results are contained 
in Sections 4 and 5; they concern the construction of coherent states of the Perelomov 
type (Section 4) and of the Barut-Girardello type (Section 5) both in finite and 
infinite dimensions. Some common properties are given in Section 6. Conclusions 
and perspectives close this paper in Section 7.

The present communication is based on an invited talk to M.R.K. given at the Physics 
Conference TIM--11 (24--26 November 2011, Timi\c soara, Romania). Thanks are due to the 
organizers for making possible this interesting pluri-disciplinary conference. 

\section{BASICS OF COHERENT STATES}

The harmonic oscillator algebra (or usual Weyl-Heisenberg algebra) is spanned by three 
linear operators, namely, an annihilation operator ($a^-$), a creation operator ($a^+$) 
and a number operator ($N = a^+ a^-$) satisfying the relations 
\begin{eqnarray}
[a^- , a^+] =  I, \quad [N, a^{-}] = - a^{-}, \quad [N, a^{+}] = + a^{+}, 
\quad a^+ = (a^-)^{\dagger}, \quad N = N^{\dagger}, 
\label{usual W-H algebra}
\end{eqnarray}
where $I$ is the identity operator. There three ways to define coherent states for the harmonic oscillator system: 
\begin{itemize}
	\item as eigenvectors $\vert z \rangle$, $z \in {\bf Z}$, of an annihilation operator 
	$a^-$ ($\Rightarrow$ Barut-Girardello type states)
	\item by acting with displacement operator $\exp(za^+ - \bar{z}a^-)$ on ground state 
	$\vert 0 \rangle$ of $N$ 
($\Rightarrow$ Perelomov type states) 
	\item by minimizing the uncertainty relation for the position and momentum operators 
	associated with $a^-$ and $a^+$ ($\Rightarrow$ Roberston-Schr\"odinger type states). 
\end{itemize}
The three ways lead to the same coherent states, a result that is not true for other dynamical 
systems. The expression for the coherent states of the harmonic oscillator, the so-called Glauber 
states [7], reads (up to a normalization factor)
$$
\vert z \rangle = \sum_{n=0}^{\infty} \frac{1}{\sqrt{n!}} z^n \vert n \rangle
$$
in terms of eigenvectors 
$$
\vert n \rangle = \frac{1}{\sqrt{n!}} (a^+)^n \vert 0 \rangle 
$$
of operator $N$. 

The situation is different for other dynamical systems or algebras. This is well-known for the 
su$(1,1)$ algebra: the Perelomov states obtained through the action of displacement operator 
$\exp(zK^+ - \bar{z}K^-)$ on ground state $\vert k, 0 \rangle$ of the positive discrete series representation 
of su$(1,1)$ are different from the Barut-Girardello states arising from eigenvalue equation 
$K_- | z \rangle = z | z \rangle$ ($K^+$ and $K_-$ are the two ladder operators of su$(1,1)$). In the 
su$(2)$ case, it is possible to define Perelomov states owing to displacement operator $\exp(zJ^+ - \bar{z}J^-)$ 
acting on ground state $\vert j , -j \rangle$ of the $2j + 1$-dimensional representation of su$(2)$ ($J^+$ and 
$J_-$ are the two ladder operators of su$(2)$); however, it is not possible to define Barut-Girardello states 
$\vert z \rangle$, $z \in {\bf C}$, for su$(2)$ as eigenstates of $J_-$. 

\section{GENERALIZED WEYL-HEISENBERG ALGEBRA} 

\subsection{Polynomial Weyl-Heisenberg algebra}

Following many works on possible extensions of the usual Weyl-Heisenberg algebras [17-27], let us consider 
the algebra spanned by an annihilation operator ($a^-$), a creation operator ($a^+$) and a number operator 
($N \not= a^+ a^-$) satisfying the commutation relations 
\begin{eqnarray}
[a^- , a^+] =  G(N), \quad [N, a^{-}] = - a^{-}, \quad [N, a^{+}] = + a^{+}, 
\label{generalized W-H algebra}
\end{eqnarray}
with
\begin{eqnarray}
a^+ = (a^-)^{\dagger}, \quad N = N^{\dagger}, \quad G(N) = F(N+1) - F(N),  
\label{hermiticity relations}
\end{eqnarray}
where the $F$ structure function is defined by 
\begin{eqnarray}
F(N) = N \prod_{i = 1}^{r} [I + \kappa_i(N-I)], \quad \kappa_i \in {\bf R} \quad (i = 1, 2, \ldots, r). 
\label{stucture function}
\end{eqnarray}
Equations (\ref{generalized W-H algebra})--(\ref{stucture function}) 
constitute a polynomial extension of the usual Weyl-Heisenberg algebra defined by (\ref{usual W-H algebra}). This polynomial Weyl-Heisenberg algebra, denoted as ${\cal A}_{\{\kappa\}}$, depends on $r$ real parameters. Of course, other choices for $F(N)$ lead to other generalized Weyl-Heisenberg algebras. 

Three interesting particular cases for $F(N)$ correspond to 
$$
\kappa_1 = \kappa, \qquad \kappa_2 = \kappa_3 = \ldots = \kappa_r = 0. 
$$
Then, the special case where $\kappa = 0$ ($F(N) = N \Rightarrow G(N) = I$) corresponds to the usual harmonic oscillator system (described by the $h_4$ usual Weyl-Heisenberg algebra). Furthermore, the cases $\kappa > 0$ and $\kappa < 0$ describe the P\"oschl-Teller system (described by the 
su$(1,1)$ algebra) and the Morse system (described by the su$(2)$ algebra), respectively [25, 27, 28]. 

\subsection{Representation of the polynomial Weyl-Heisenberg algebra}

Going back to the general case, since the ${\cal A}_{\{\kappa\}}$ algebra is an extension of the usual oscillator algebra, we may hope to find a representation of ${\cal A}_{\{\kappa\}}$ which extends that of $h_4$. Indeed, it is easy to check that the actions 
	\begin{eqnarray}
	&& a^-\vert n \rangle = \sqrt{F(n)}   e^{{+i [F(n) - F(n-1)]\varphi}} \vert n - 1 \rangle, \quad a^-\vert 0 \rangle = 0,
	\label{rep1}  \\ 
	&& a^+\vert n \rangle = \sqrt{F(n+1)} e^{{-i [F(n+1) - F(n)]\varphi}} \vert n + 1 \rangle, \quad N \vert n \rangle = n \vert n \rangle 
	\label{rep3} 
	\end{eqnarray}
(on the Hilbert space spanned by the eigenvectors of $N$) formally define a representation of ${\cal A}_{\{\kappa\}}$. The $\varphi$ parameter is a real parameter which is generally taken to be 0 in 
developments concerning the harmonic oscillator; we shall see that 
this parameter is essential to ensure temporal stability of coherent states. Note that 
$$
a^+ a^- = F(N),  
$$
a relation that generalizes $N = a^+ a^-$ for the harmonic oscillator 
and gives a significance to the $F$ function: $F(N)$ can be considered as the Hamiltonian for a quantum system.  

We may now ask what is the dimension of the representation (Fock-Hilbert) space generated by the orthonormal 
set $\{ | n \rangle : n \ {\rm ranging} \}$? The dimension of the representation of ${\cal A}_{\{\kappa\}}$ 
afforded by (\ref{rep1}) and (\ref{rep3}) is controlled by the positiveness of:
$$
F(n) = n \prod_{i = 1}^{r} [1 + \kappa_i(n-1)] \geq 0. 
$$
We shall limit ourselves here to two cases. 
\begin{itemize}
	\item $\kappa_i \geq 0$ ($i= 1, 2, \ldots, r$): there is no limit to the number of states $\vert n \rangle$ and the representation is infinite-dimensional so that the Fock-Hilbert space is generated by $\{ | n \rangle : n \in {\bf N} \}$, 
	\item $\kappa_1 < 0$, $\kappa_i \geq 0$ ($i= 2, 3, \ldots, r$): the  number of states $\vert n \rangle$ is limited and the representation has dimension $d$ with 
	\begin{eqnarray}
	d = 1 - \frac{1}{\kappa_1}, \quad -1/\kappa_1 \in {\bf N}^*
\quad \Rightarrow \quad F(n) = n \frac{d-n}{d-1} \prod_{i = 2}^{r} [1 + \kappa_i(n-1)], 
	\nonumber
	\end{eqnarray}
so that the Fock-Hilbert space is generated by $\{ | n \rangle : n = 0, 1, \ldots, d-1 \}$.
	\end{itemize}

In the finite-dimensional case, two further conditions are verified. Indeed, it can be shown that  
	$$
	a^+ \vert d-1 \rangle = 0, \quad 
	(a^-)^{d} = (a^+)^{d} = 0, 
	$$
two relations that generalize the conditions for $k$-fermions [11, 12] (the 
$d=k=2$ case corresponds to ordinary fermions).

\subsection{Truncated polynomial Weyl-Heisenberg algebra}

In the infinite-dimensional case, it can be useful to truncate the representation space to a subspace of dimension $s$ (for defining a unitary phase operator or for perturbation theory purposes). This can be achieved via the Pegg-Barnett trick developed for the $h_4$ oscillator algebra [29]. This amounts to replace the $a^{\pm}$ operators by 
  $$
 	a^{\pm}(s)= a^{\pm}   - \sum_{n=s}^{\infty} \sqrt{F(n)} e^{{\mp}i [F(n)-F(n-1)]\varphi} \vert n - \frac{1}{2} \pm \frac{1}{2} \rangle \langle n - \frac{1}{2} \mp \frac{1}{2} \vert. 
  $$
Therefore, we pass from the ${\cal A}_{\{\kappa\}}$ algebra to the ${\cal A}_{\{\kappa,s\}}$ truncated algebra defined by 
	$$
	[a^-(s) , a^+(s)] = G_s(N) - F(s) \vert s-1 \rangle \langle s-1 \vert, \quad 
	[N , a^{\pm}(s)] = \pm a^{\pm}(s), 
	$$
with
	$$
a^+(s) = (a^-(s))^{\dagger}, \quad N = N^{\dagger}, \quad	G_s(N) = \sum_{n=0}^{s-1} [F(n+1)-F(n)] \vert n \rangle \langle n \vert.
	$$
Thus, the results derived for a Weyl-Heisenberg algebra with a representation of dimension $d$ can be applied to a ${\cal A}_{\{\kappa,s\}}$ truncated algebra arising from another  Weyl-Heisenberg algebra with an infinite-dimensional representation.

\section{PERELOMOV TYPE COHERENT STATES}

The derivation of {\it \`a la} Perelomov coherent states for an arbitrary ${\cal A}_{\{\kappa\}}$ algebra from the action of a displacement 
operator on state $\vert 0 \rangle$ is very difficult because commutator $[a^- , a^+]$ differs from the identity operator. Consequently, we 
shall adopt a more simple strategy based on the use of a Fock-Bargmann space associated with ${\cal A}_{\{\kappa\}}$. This strategy can be 
summed up as follows. 

Let us look for states in the form  
	\begin{eqnarray}
	\vert z , \varphi \rangle =  \sum_{n} {\overline{a_n}} z^n \vert n \rangle, \quad a_n \in {\bf C}, \quad z \in {\bf C}, 
	\label{forme des ec}
	\end{eqnarray}
where the sum on $n$ is finite or infinite according to as ${\cal A}_{\{\kappa\}}$ admits a finite- or infinite-dimensional representation. The 
$a_n$ coefficients can then be determined from the correspondence rules 
	\begin{eqnarray}
	\vert n \rangle \longrightarrow a_{n} z^{n}, \qquad
	a^-\longrightarrow \frac{d}{dz}
	\label{correspondance pour Perelomov}
	\end{eqnarray}
applied to relations (\ref{rep1}) and (\ref{rep3}). The convergence of the $\vert z , \varphi \rangle$ states so-obtained should be checked as well as their 
existence as Perelomov type coherent states.

\subsection{The infinite case}

The strategy just described leads to the following recurrence relation
	\begin{eqnarray}
	n a_{n}= \sqrt{F(n)}e^{+i[F(n)-F(n-1)]\varphi}a_{n-1}, 
	\label{recur}
	\end{eqnarray}
which can be iterated to give 
	\begin{eqnarray}
	a_{n}= \frac{\sqrt{F(n)!}}{n!}e^{+iF(n)\varphi}, 
	\label{recur2}
	\end{eqnarray}
(by taking $a_0 = 1$). In Eq.~(\ref{recur2}), the generalized factorials 
are defined by 
$$
F(0)! = 1, \qquad F(n)! = F(1) F(2) \ldots F(n). 
$$
This yields the following result.

{\bf Result 1}. In infinite dimension, the states
  $$
	\vert z , \varphi \rangle = \sum_{n=0}^{\infty} 
  \frac{\sqrt{F(n)!}}{n!} z^n e^{-iF(n)\varphi} \vert n \rangle
	$$
exist only for $r=1$ in the disk 
$\{ z \in {\bf C} \ : \ \vert z \vert < {1}/{\sqrt{\kappa_1}} \}$. They 
satisfy $\vert z , \varphi \rangle = \exp( z a^+) \vert 0 \rangle$ 
and are thus coherent states in the Perelomov sense. 

We note that the restriction on $r$ comes from the fact that the $\vert z , \varphi \rangle$ 
states cannot be normalized if $r \geq 2$. 

{\bf Example 1}. Let us examine the case where 
$r = 1$ and $\kappa_1 = {1}/{\ell}$ with $\ell \in {\bf N}^*$.
The corresponding coherent states read
$$
	\vert z , \varphi \rangle = \sum_{n = 0}^{\infty}
	\sqrt{ \frac{1}{n!} \frac{(\ell - 1 + n)!}{\ell^n (\ell-1)!} } z^n e^{-iF(n)\varphi}
	\vert n \rangle.  
$$
Note that the $\ell \to \infty$ limit corresponds to the harmonic oscillator. 

\subsection{The finite case}

In this case, recurrence relation (\ref{recur}) is valid. However, there is no restriction on $r$ 
for normalization purposes. We are thus left with Result 2. 

{\bf Result 2}. In finite dimension ($dim = d$ or $s$), the states
  $$
	\vert z , \varphi \rangle = \sum_{n=0}^{dim - 1} 
  \frac{\sqrt{F(n)!}}{n!} z^n e^{-iF(n)\varphi} \vert n \rangle
	$$
exist for any value of $r$ and any $z$ in ${\bf C}$. They satisfy $\vert z , \varphi \rangle = \exp( z a^+) \vert 0 \rangle$ 
and are thus coherent states in the Perelomov sense.

{\bf Example 2}. For $r = 1$ and $dim = d$ (the 
${\cal A}_{\{\kappa\}}$ algebra has a representation of dimension $d$), 
$\vert {{z}} , \varphi \rangle$ reads
$$
	\vert {{z}} , \varphi \rangle = \sum_{n = 0}^{d-1} 
	\sqrt{ \frac{1}{n!} \frac{(d-1)!}{(d-1)^n (d-1-n)!} } {{z}}^n
	e^{-iF(n)\varphi}\vert n \rangle. 
$$
Note that the $d \to \infty$ limit corresponds to the harmonic oscillator.

\section{BARUT-GIRARDELLO TYPE COHERENT STATES}

A strategy similar to that used for coherent states of the Perelomov type can be set up for the determination of Barut-Girardello type 
coherent states associated with ${\cal A}_{\{\kappa\}}$. It consists in looking for states in the form given by (\ref{forme des ec}) 
and in replacing (\ref{correspondance pour Perelomov}) by 
	\begin{eqnarray}
	\vert n \rangle \longrightarrow a_{n} z^{n}, \qquad	a^+ \longrightarrow {z}.	
	\label{correspondance pour B-G}
	\end{eqnarray}
 
\subsection{The infinite case}

By introducing Eq.~(\ref{correspondance pour B-G}) in (\ref{rep1}) and (\ref{rep3}), we get the recurrence relation
	$$
	a_{n}= \sqrt{F(n+1)}e^{-i[F(n+1)-F(n)]\varphi} a_{n+1}
	$$
which admits the solution 
  $$
	a_{n}= \frac{1}{\sqrt{F(n)!}}e^{+iF(n)\varphi}
	$$
(we take $a_0 = 1$). As a conclusion, we have the next result.

{\bf Result 3}. In infinite dimension, the states
  $$
	\vert z , \varphi \rangle = \sum_{n=0}^{\infty} 
  \frac{1}{\sqrt{F(n)!}} z^n e^{-iF(n)\varphi} \vert n \rangle
	$$
exist for any value of $r$ in the whole complex plane ${\bf C}$. They 
satisfy $a^- \vert z , \varphi \rangle = z \vert z , \varphi \rangle$ 
and are thus coherent states in the Barut-Girardello sense. 

{\bf Example 3}. In the special case where 
$r = 1$ and $\kappa_1 = {1}/{\ell}$ with $\ell \in {\bf N}^*$, 
we have
$$
	\vert z , \varphi \rangle = \sum_{n = 0}^{\infty}
	\sqrt{ \frac{1}{n!} \frac{\ell^n (\ell-1)!}{(\ell - 1 + n)!} } z^n e^{-iF(n)\varphi}
	\vert n \rangle 
$$
Note that the $\ell \to \infty$ limit corresponds to the harmonic oscillator.

\subsection{The finite case}

The situation is quite new in finite dimension ($dim = d$ or $s$). Indeed, the strategy applied in the last subsection to the infinite case requires that either the $\vert {{z}} , \varphi \rangle$ states are identically 0 or $z^{dim} = 0$. Therefore, there is only the trivial solution if $z$ is a complex variable. However, if $z$ is replaced by a Grassmann variable, $\theta$, of order $dim$ (i.e., 
$\theta^{dim} = 0$), we obtain the following result. 

{\bf Result 4}. In finite dimension ($dim = d$ or $s$), there are no Barut-Girardello coherent states for $z \in {\bf C}$. However, Barut-Girardello coherent states exist for
  $$
	z \to \theta = \ {\rm {Grassmann} \ {variable} \ with} \ \theta^{dim} = 0. 
	$$
They are given by
  $$
	\vert \theta , \varphi \rangle = \sum_{n=0}^{dim-1} 
  \frac{1}{\sqrt{F(n)!}} \theta^n e^{-iF(n)\varphi} \vert n \rangle
	$$	
for any value of $r$ and satisfy $a^- \vert \theta , \varphi \rangle = \theta \vert \theta , \varphi \rangle$. 

It should be noted that when $dim \to \infty$ and $\theta \to z$, we get back the coherent states for the harmonic oscillator. 

{\bf Example 4}. For $r = 1$ and $dim = d = 2$, we have the states 
	$$
	\vert \theta , \varphi \rangle = 
  \vert 0 \rangle + \theta e^{-i \varphi} \vert 1 \rangle, 
	$$
which for $\varphi = 0$ coincide with the coherent states for 
the fermionic oscillator [30] (of interest for qubits).

\section{COMMON PROPERTIES}

The Perelomov and Barut-Girardello coherent states derived above share some common properties 
which can be summarized as follows. (Further details shall be published elsewhere [31].)
\begin{itemize}
	\item They are continuous in the variables $\varphi$ and $z$ or $\theta$.
	\item They are stable under time evolution, i.e., 
$$
e^{-i H t} | z \ {\rm or} \ \theta, \varphi \rangle = | z \ {\rm or} \ \theta, \varphi + t \rangle, \quad H = F(N) = a^+ a^-.
$$
	\item They are normalizable but not orthogonal. 
	\item They satisfy overcompleteness relations, i.e.,  
$$
	\int d\mu (\vert z \vert) \vert z , \varphi \rangle \langle z , \varphi \vert = \sum_{n=0}^{\infty} \vert n \rangle \langle n \vert \ {\rm or} \ 
	\int d\mu (\vert z \vert) \vert z \ {\rm or} \ \theta , \varphi \rangle \langle z \ {\rm or} \ \theta , \varphi \vert = \sum_{n=0}^{dim -1} \vert n \rangle \langle n \vert
$$
in infinite or finite dimension, where the $d\mu$ measures are given in [31] in terms of special functions (complex variable case) 
or generalized Berezin calculus (Grassmann variable case). 
\end{itemize}

\section{SUMMARY, CONCLUSIONS AND PERSPECTIVES}

We focused in this work on a $r$-parameter polynomial Weyl-Heisenberg algebra, ${\cal A}_{\{\kappa\}}$, 
that generalizes the oscillator algebra. We showed that this algebra admits infinite- or finite-dimensional 
representations depending on the value of the parameters. In addition, ${\cal A}_{\{\kappa\}}$can describe   
dynamical quantum systems with nonlinear (in $n$) spectra and can serve as a framework for generating phase 
operators, phase states and mutually unbiased bases (for $r=1$, see [27, 32]). We developed a  simple and 
straightforward derivation, in a Fock-Bargmann approach, of Perelomov and Barut-Girardello coherent states 
for finite- and infinite-dimensional representations of ${\cal A}_{\{\kappa\}}$. It is to be noted that our 
construction of Barut-Girardello coherent states in dimension $d$ in terms of Grassmann variables establishes 
a link with $k$-fermions [11, 12] which are objects interpolating between fermions ($k=d=2$) and bosons 
($k=d \to \infty$). 

As open questions and perspectives, we can mention: a probabilistic interpretation 
and the study of Bargmann functions associated with some of the coherent states 
obtained in this work. In this respect, we close with some preliminary developments 
concerning Bargmann functions for the Barut-Girardello type coherent states in infinite 
dimension. 

We shall restrict ourselves to the case
$$
\kappa_i = {1}/{\ell}_i, \quad \ell_i \in {\bf N}^* \ (i = 1, 2, \ldots, r).
$$
Then, the Barut-Girardello coherent states given in Result 3 can be normalized as
  $$
	\vert z , \varphi \rangle = {\cal N}^{-1} \sum_{n=0}^{\infty} 
  \frac{1}{\sqrt{F(n)!}} z^n e^{-iF(n)\varphi} \vert n \rangle, \quad 
  \vert {\cal N} \vert^{2} = {}_0F_r(\ell_1, \ell_2, \ldots, \ell_r; ~\ell_1 \ell_2 \ldots \ell_r \vert z \vert^2). 
	$$
With respect to these coherent states, the vector 
$$
\vert f \rangle = \sum_{n=0}^{\infty} f_n \vert n \rangle, \quad \sum_{n=0}^{\infty} \vert f_n \vert^2 < \infty
$$
can be represented by the $f_{\varphi}$ analytical function defined by 
$$ 
f_{\varphi}(z) = \sum_{n=0}^{\infty} 
\frac{1}{\sqrt{F(n)!}} z^n e^{-iF(n)\varphi} f_n. 
$$
Let us recall that the growth of an arbitrary entire series, say $f(z) = \sum_{n=0}^{\infty} c_n z^n$, 
is described by means of two nonnegative numbers: order $\rho$ and type $\sigma$ given by [33]
$$
\rho = \lim_{n \to \infty} \bigg( - n \frac{\log n}{\log \vert c_n \vert} \bigg), \quad 
\sigma  =  \frac{1}{e\rho} \lim_{n\to \infty} \bigg(n \vert c_n \vert^{\frac{\rho}{n}}\bigg). 
$$
This allows to classify entire functions according to their growth as $\vert z \vert \to \infty$: 
the maximum modulus $M(R)$ of $f(z)$ for $\vert z \vert = R$ behaves like
$$ 
M(R) \sim \exp( \sigma  \vert z \vert^{\rho} )
$$
as $R$ goes to infinity. It is simple to verify (through the use of Schwarz inequality) that
$$
\vert f_{\varphi} (z) \vert \leq \vert {\cal N} \vert
$$
and, using arguments similar to those in [9], it can be shown that order $\rho$ and type $\sigma$ 
of the $f_{\varphi}$ function are  
	\begin{eqnarray}
\rho = \frac{2}{1+r}, \quad \sigma = \frac{1+r}{2} (\ell_1 \ell_2 \ldots \ell_r)^{\frac{1}{1+r}}. 
	\label{comparison with Vourdas}
	\end{eqnarray}
It is interesting to note that the $\rho$ order of the Bargmann functions
associated with Barut-Girardello coherent states decreases as $r$ increases. In the particular case 
where $r = 1$ and $\ell_1 = 1$, Eq.~(\ref{comparison with Vourdas}) is in agreement with the result 
for Example B of [34] which corresponds in our notation to $F(n)! = (n!)^2$. (Of course, the standard 
harmonic oscillator case is trivial and can be recovered by setting $F(n) = n$.) Finally, let us mention 
that Eq.~(\ref{comparison with Vourdas}) can also be derived from the behavior of the measure for 
the Barut-Girardello states (see [31]) by using a method similar to that of [34].

%%%%%%%%%%%%%%%%%%%%%%%%%%%%%%%%%%%%%%%%%%%%%%%%
%% BACKMATTER
%%%%%%%%%%%%%%%%%%%%%%%%%%%%%%%%%%%%%%%%%%%%%%%%

%%% \begin{theacknowledgments}
%%% \end{theacknowledgments}

%%%%%%%%%%%%%%%%%%%%%%%%%%%%%%%%%%%%%%%%%%%
%% The following lines show an example how to produce a bibliography
%% without the help of the BibTeX program. This could be used instead
%% of the above.
%%%%%%%%%%%%%%%%%%%%%%%%%%%%%%%%%%%%%%%%%%%

\end{document}